\documentstyle[psfig,a4,11pt,thmsa,sw20lart]{article}
\input tcilatex
\begin{document}

\topmargin 0pt \oddsidemargin 5mm

\setcounter{page}{1}

\hspace{8cm}{} \vspace{2cm}

\begin{center}
{\large {THE DYNAMICS OF CHARGES INDUCED BY A CHARGED PARTICLE TRAVERSING A
DIELECTRIC SLAB}}\\

{H.B. Nersisyan}\\\vspace{1cm} {\em Division of Theoretical Physics,
Institute of Radiophysics and Electronics, Alikhanian Brothers St. 2,
Ashtarak-2, 378410, Republic of Armenia}\footnote{%
E-mail: Hrachya@irphe.sci.am}
\end{center}

\vspace {5mm} \centerline{{\bf{Abstract}}}

We studied the dynamics of surfacea and wake charges induced by a charged
particle traversing a dielectric slab. It is shown that after the crossing
of the slab first boundary, the induced on the slab surface charge (image
charge) is transformed into the wake charge, which overflows to the second
boundary when the particle crosses it. It is also shown, that the
polarization of the slab is of an oscillatory nature, and the net induced
charge in a slab remains zero at all stages of the motion.

\newpage 

\section{Introduction}

As it passes through a medium, a fast charged particle excites oscillations
of the charge density behind itself [1-3]. These wakefields and the particle
energy losses associated with their excitation have been studied widely for
a variety of media [3-8]. Wakefields have recently reattracted interest
because of the development of new methods for accelerating particles [9-11].

In most studies of wakefields it has been assumed that the medium is
unbounded. The wakefields are excited as the particle enters the medium, or
they disappear when the particle leaves the medium, because of various
transient polarization processes which occur near the interface. Among these
processes, the excitation of surface oscillations and the associated
additional energy loss have been studied previously [12-20]. In connection
with the development of new particle acceleration methods, numerical
calculations have determined the distance from the sharp plasma boundary at
which the amplitude of the wakefield excited by an ultrarelativistic
particle reaches the same level as in an unbounded medium [21].

Fairly recently, in connection with problems of emission electronics and
optoelectronics, the image charge and the dynamical image potential created
by a moving particle has also been investigated. To describe the process of
formation of the image charge, various approaches (quantum mechanical, the
hydrodynamic, etc.) and various models of the medium have been employed
[22-25].

In the present paper we analyze the dynamics of reversal of the sign of the
charges induced at the slab boundary (repolarization of the slab) as the
particle crosses the interface. The process is found to be of a
nonmonotonic, oscillatory nature. The case of normal incidence of a particle
through the slab is considered.

The paper outline is as follows. In section 2, general expressions for the
density of wake charge and total wake charge have been found, using
Poisson's equation. In section 3, general expressions for the density of
induced surface charges and total charges have been found, using expressions
for the normal component of the electric field in the internal and external
space of the slab [26]. We apply the results obtained in sections 2 and 3 to
the case when slab constructed from a diatomic cubic ionic crystal or polar
semiconductor. In section 4, the obtained results are discussed.

\section{The electromagnetic field of charged particle traversing a slab}

We consider a fast particle of charge $q$ moving with a velocity $u$ along
the $z$-axis normal to the boundaries of a slab characterized by a local
dielectric function $\varepsilon (\omega )$. The time interval $t$ during
which the particle moves through the medium is $0<t<a/u$, where $a$ is the
slab thickness. Outside this interval the particle moves in a vacuum.

Ginzburg and Tsytovich [26] have given the expressions for the
electromagnetic field of a fast charge passing through a slab. We shall
briefly repeat the method of obtaining of these expressions.

Since the problem is homogeneous both in time and directions in each domain $%
z<0$, $0<z<a$, and $z>a$ normal to the charge velocity, it is convenient to
represent all field components as Fourier integrals over time and
transversal coordinates ${\bf r}=(x,y)$. Then the Fourier component of
electric field is obtained from the Maxwell equations:

\begin{equation}
\left[ \frac{\partial ^2}{\partial z^2}+\frac{\omega ^2}{c^2}\varepsilon
(\omega )-k^2\right] {\bf E}({\bf k},\omega ,z)=4\pi \left[ -\frac{i\omega u%
}{c^2}{\bf n}+\frac 1{\varepsilon (\omega )}\left( i{\bf k}+{\bf n}\frac
\partial {\partial z}\right) \right] \rho _0({\bf k},\omega ,z),
\end{equation}
where ${\bf n}={\bf u}/u$, ${\bf k}=(k_x,k_y)$, $\rho _0({\bf k},\omega ,z)$
is the Fourier component of the charge density of the particle

\begin{equation}
\rho _0({\bf k},\omega ,z)=\frac q{(2\pi )^3u}\exp \left( i\frac \omega
uz\right) .
\end{equation}
The Fourier component of the magnetic field is expressed through ${\bf E}(%
{\bf k},\omega ,z)$ as follows:

\begin{equation}
{\bf B}({\bf k},\omega ,z)=\frac c\omega \left\{ -i\left[ \nabla \times {\bf %
E}({\bf k},\omega ,z)\right] +\left[ {\bf k}\times {\bf E}({\bf k},\omega
,z)\right] \right\} .
\end{equation}

The total solution of (1) for a charge density (2) is a sum of solutions to
homogeneous and inhomogeneous equations. While the first equation describes
the radiation field, the second equation describes the particle field proper
in a medium with local dielectric function $\varepsilon (\omega )$. Also,
equation (1) must be solved for each domain inside and outside the slab, and
therefore the solutions are joined using the boundary conditions (equality
of normal induction components and transverse electric field components on
the boundary)

\begin{eqnarray}
E_z({\bf k},\omega ,-0) &=&\varepsilon (\omega )E_z({\bf k},\omega
,+0),\quad \varepsilon (\omega )E_z({\bf k},\omega ,a-0)=E_z({\bf k},\omega
,a+0), \\
{\bf k\cdot E}({\bf k},\omega ,-0) &=&{\bf k\cdot E}({\bf k},\omega
,+0),\quad {\bf k\cdot E}({\bf k},\omega ,a-0)={\bf k\cdot E}({\bf k},\omega
,a+0).  \nonumber
\end{eqnarray}

Taking account of these conditions, the following system of relations is
obtained:

\begin{eqnarray}
{\bf E}({\bf k},\omega ,z) &=&{\bf E}^{(1)}({\bf k},\omega )\exp \left(
i\frac \omega uz\right) +\frac{2iq}{(2\pi )^2k\omega }a_1^{(-)}\left( k{\bf n%
}+{\bf k}\frac \omega {kc}\tau _1\right) \exp \left( -i\frac \omega c\tau
_1z\right) , \\
z &<&0  \nonumber
\end{eqnarray}

\begin{eqnarray}
{\bf E}({\bf k},\omega ,z) &=&{\bf E}^{(2)}({\bf k},\omega )\exp \left(
i\frac \omega uz\right) +\frac{2iq}{(2\pi )^2k\omega }\left[ a_2^{(-)}\left(
k{\bf n}+{\bf k}\frac \omega {kc}\tau _2\right) \exp \left( -i\frac \omega
c\tau _2z\right) +\right. \\
&&\left. +a_2^{(+)}\left( k{\bf n-k}\frac \omega {kc}\tau _2\right) \exp
\left( i\frac \omega c\tau _2z\right) \right] ,  \nonumber \\
0 &\leq &z\leq a  \nonumber
\end{eqnarray}

\begin{eqnarray}
{\bf E}({\bf k},\omega ,z) &=&{\bf E}^{(1)}({\bf k},\omega )\exp \left(
i\frac \omega uz\right) +\frac{2iq}{(2\pi )^2k\omega }a_1^{(+)}\left( k{\bf %
n-k}\frac \omega {kc}\tau _1\right) \exp \left( i\frac \omega c\tau
_1z\right) , \\
z &>&a  \nonumber
\end{eqnarray}
where 
\begin{equation}
{\bf E}^{(1)}({\bf k},\omega )=-\frac{2iq}{(2\pi )^2}\frac{\omega {\bf n}%
+\gamma ^2u{\bf k}}{\omega ^2+\gamma ^2k^2u^2},
\end{equation}

\begin{equation}
{\bf E}^{(2)}({\bf k},\omega )=-\frac{2iq}{(2\pi )^2}\frac{\omega \left[
1-\beta ^2\varepsilon (\omega )\right] {\bf n}+u{\bf k}}{\varepsilon (\omega
)\left\{ \omega ^2\left[ 1-\beta ^2\varepsilon (\omega )\right]
+k^2u^2\right\} },
\end{equation}

\begin{equation}
\tau _1=\sqrt{1-k^2c^2/\omega ^2},\quad \tau _2=\sqrt{\varepsilon (\omega
)-k^2c^2/\omega ^2},
\end{equation}

\begin{eqnarray}
a_1^{(-)} &=&-\frac{\beta k^2c^2/\omega ^2}{\left( 1-\beta ^2\tau
_1^2\right) \left( 1-\beta ^2\tau _2^2\right) }\frac{1-\varepsilon (\omega )%
}{D(k,\omega )}\times \\
&&\times \left\{ f_{-}^{(1)}\exp \left( i\frac \omega c\tau _2a\right)
+f_{-}^{(2)}\exp \left( -i\frac \omega c\tau _2a\right) +f_{-}^{(3)}\exp
\left( i\frac \omega ua\right) \right\} ,  \nonumber
\end{eqnarray}

\begin{eqnarray}
a_1^{(+)} &=&\frac{\beta k^2c^2/\omega ^2}{\left( 1-\beta ^2\tau _1^2\right)
\left( 1-\beta ^2\tau _2^2\right) }\frac{1-\varepsilon (\omega )}{D(k,\omega
)}\exp \left[ i\frac \omega ua\left( 1-\beta \tau _1\right) \right] \times \\
&&\times \left\{ f_{+}^{(1)}\exp \left( i\frac \omega c\tau _2a\right)
+f_{+}^{(2)}\exp \left( -i\frac \omega c\tau _2a\right) +f_{+}^{(3)}\exp
\left( -i\frac \omega ua\right) \right\} ,  \nonumber
\end{eqnarray}

\begin{equation}
f_{\pm }^{(1)}=\left( \tau _2-\varepsilon \tau _1\right) \left( 1\mp \beta
\tau _2\right) \left( 1\pm \beta \tau _2-\beta ^2\right) ,
\end{equation}

\begin{equation}
f_{\pm }^{(2)}=\left( \tau _2+\varepsilon \tau _1\right) \left( 1\pm \beta
\tau _2\right) \left( 1\mp \beta \tau _2-\beta ^2\right) ,
\end{equation}

\begin{equation}
f_{\pm }^{(3)}=2\tau _2\left[ \beta ^2\left( 1+\varepsilon -k^2c^2/\omega
^2\right) -1\pm \beta ^2\varepsilon \tau _1\right] ,
\end{equation}

\begin{equation}
D(k,\omega )=\left( \tau _2+\varepsilon \tau _1\right) ^2\exp \left( -i\frac
\omega c\tau _2a\right) -\left( \tau _2-\varepsilon \tau _1\right) ^2\exp
\left( i\frac \omega c\tau _2a\right)
\end{equation}
and $\beta =u/c$, $\gamma ^{-2}=1-\beta ^2$. The functions $a_2^{(-)}$ and $%
a_2^{(+)}$ are expressed through $a_1^{(-)}$ and $a_1^{(+)}$ and are not
explicitly given here. They may be obtained from the matching conditions for
the normal component of the electric induction on the surfaces $z=0$ and $%
z=a $.

The first terms in (6) and (8) describe the Coulomb field of the particle.
The first term in (7) describes the particle field in an unbounded medium
characterized by the dielectric function $\varepsilon (\omega )$. The field
is identical with a Cherenkov radiation electric field in the frequency
range $\beta ^2\varepsilon (\omega )>1$. All other terms are due to
existence of boundaries. Particularly, they describe the transition
radiation in the backward (second term in (6)) and forward (second term in
(8)) directions [26].

\section{The wake charge evaluation}

In this section we shall consider the volume charge induced by a moving
particle in a slab (the so-called wake charge). To evaluate the wake charge,
Poisson equation is used

\begin{equation}
\rho _v=(1/4\pi )\nabla {\bf E}-\rho _0
\end{equation}
in which the $\rho _0=q\delta ({\bf r})\delta (\xi )$ is the charge density
of a test particle, $\xi =z-ut$, ${\bf E}$ is the electric field in the slab
which is determined by the inverse Fourier transformation of (7). Since the
divergence of the second term in (7) (of the radiation field) is zero, the
wake-charge density is determined only by the first term in (7).

Using this term in the relation (18) we obtain

\begin{equation}
\rho _v=\frac q{2\pi u}\delta ({\bf r})\int_{-\infty }^{+\infty }d\omega
\exp \left( i\frac \omega u\xi \right) \frac{1-\varepsilon (\omega )}{%
\varepsilon (\omega )},
\end{equation}
where $\delta (x)$ is a Dirac function. Since the dielectric function of the
medium has poles only in the lower $\omega $ half-plane [6], no induced
charge exists in front ($\xi >0$) of the particle. Note that the relation
(19) may be also obtained from the expression for electrostatic potential
created by the particle in unbounded medium described by a dielectric
function $\varepsilon (\omega )$ [6].

Evaluating an integral of expression (19) over the volume we obtain the wake
charge, induced by the particle, moving in the slab:

\begin{equation}
Q_v(t)=-q\left[ \Phi (t)-\Phi (\tau )\right] ,
\end{equation}
where 
\begin{eqnarray}
\Phi (t) &=&\frac 1{2\pi i}P\int_{-\infty }^{+\infty }\frac{d\omega }\omega
\exp \left( -i\omega t\right) \frac{1-\varepsilon (\omega )}{\varepsilon
(\omega )}= \\
&=&-\frac 12\left( 1-\frac 1{\varepsilon _0}\right) +\theta (t)\left\{
1-\frac 1{\varepsilon _0}-\sum_j\exp \left( -\upsilon _jt\right) \left[
A_j\cos \left( \omega _jt\right) +B_j\sin \left( \omega _jt\right) \right]
\right\}  \nonumber
\end{eqnarray}
and $\tau =t-a/u$, $\varepsilon _0$ is the static dielectric constant of the
medium, $\theta (t)$ is the Heaviside unit step function (with $\theta
(0)=\frac 12$), the symbol $P$ denotes the principal value of the integral, $%
\pm \omega _j-i\upsilon _j$ are the solutions of the equation $\varepsilon
(\omega )=0$ ($\upsilon _j>0$), while coefficients $A_j$ and $B_j$ are

\begin{equation}
A_j=2{\rm Re}\left\{ \frac 1{\left( \omega _j-i\upsilon _j\right)
\varepsilon ^{^{\prime }}\left( \omega _j-i\upsilon _j\right) }\right\}
,\quad B_j=2{\rm Im}\left\{ \frac 1{\left( \omega _j-i\upsilon _j\right)
\varepsilon ^{^{\prime }}\left( \omega _j-i\upsilon _j\right) }\right\} .
\end{equation}
Here the prime denotes differentiation with respect to the argument. The
summation in (21) is carried over all zeros of the dielectric function.

Analytic properties of the dielectric function [6] and the residue theorem
were used in evaluation of $\Phi (t)$.

\section{Calculation of the induced surface charges}

The surface-induced charge density is related to the discontinuity of the
electric field $z$-component. Expressions (6)-(8) give 
\begin{equation}
\sigma _i({\bf r},t)=\int d^2{\bf k}\int_{-\infty }^{+\infty }d\omega \sigma
_i({\bf k},\omega )\exp [i({\bf kr}-\omega t)],
\end{equation}
where 
\begin{equation}
(2\pi )^2\sigma _0({\bf k},\omega )=\frac{iq}{2\pi \omega }\frac{%
1-\varepsilon (\omega )}{\varepsilon (\omega )}\left[ -\frac{\omega ^2}{%
\omega ^2+\gamma ^2k^2u^2}+a_1^{(-)}\right] ,
\end{equation}

\begin{equation}
(2\pi )^2\sigma _a({\bf k},\omega )=\frac{iq}{2\pi \omega }\frac{\varepsilon
(\omega )-1}{\varepsilon (\omega )}\left[ -\frac{\omega ^2\exp (i\omega a/u)%
}{\omega ^2+\gamma ^2k^2u^2}+a_1^{(+)}\exp [i(\omega /c)\tau _1a]\right]
\end{equation}
and indices $i=0,$ $a$ refer to the first and second boundaries.

The total induced charge is obtained by integration of expressions
(23)-(25): 
\begin{equation}
Q_{is}(t)=\int d^2{\bf r}\sigma _i({\bf r},t)=(2\pi )^2\int_{-\infty
}^{+\infty }d\omega \sigma _i(\omega )\exp (-i\omega t),
\end{equation}
where $\sigma _i(\omega )=\sigma _i({\bf k}=0,\omega )$.

Let us consider the dynamics of induced charges for the dielectric slab, the
dielectric function of which, as it is known, has no singularity in the
static limit, when $\omega \rightarrow 0$ [6]. It is clear from expressions
(12) and (13) that the functions $a_1^{(-)}$ and $a_1^{(+)}$ in expressions
(24) and (25) are proportional to $k^2$, and calculating the function $%
\sigma _i(\omega )$ for the dielectric slab the above-mentioned functions
tend to zero. Thus, the charges being induced on the surfaces of the
dielectric slab are determined only by the first terms in the expressions
(24) and (25), that is by the electric fields which are created by the
particle in an unbounded dielectric with the dielectric function $%
\varepsilon (\omega )$ and in the vacuum when the dielectric is absent.

Using the known Sokhotsky-Plemel relations [6] for quantities $\sigma
_i(\omega )$ in (26) we have

\begin{equation}
(2\pi )^2\sigma _0(\omega )=\frac q{2\pi i}\frac{1-\varepsilon (\omega )}{%
\varepsilon (\omega )}P\frac 1\omega ,
\end{equation}

\begin{equation}
(2\pi )^2\sigma _a(\omega )=-\frac q{2\pi i}\frac{1-\varepsilon (\omega )}{%
\varepsilon (\omega )}\exp \left( i\frac \omega ua\right) P\frac 1\omega .
\end{equation}
The following relations are obtained from expressions (21), (26)-(28)

\begin{equation}
Q_{os}(t)=q\Phi (t),\quad Q_{as}(t)=-q\Phi (\tau ).
\end{equation}
Thus, expressions (21) and (29) make it possible to obtain the total surface
charge as soon as the zeros of the function $\varepsilon (\omega )$ are
known. One may easily verify that, at any time, the net induced charge (a
sum of $Q_{os}$, $Q_{as}$ and $Q_v$) in the slab is zero.

The following interpretation for expressions (20), (21) and (29) may be
given. When the particle approaches the slab surface from a vacuum ($t<0$)
we have $Q_{os}=-(q/2)(1-1/\varepsilon _0)$, $Q_{as}=(q/2)(1-1/\varepsilon
_0)$, and $Q_v=0$. Note that the first boundary of the slab is charged
oppositely to the second boundary.

While the particle moves inside the medium ($0<t<a/u$), the first boundary
charge oscillates and decreases:

\begin{equation}
Q_{os}(t)=\frac q2\left( 1-\frac 1{\varepsilon _0}\right) -q\sum_j\exp
\left( -\upsilon _jt\right) \left[ A_j\cos \left( \omega _jt\right) +B_j\sin
\left( \omega _jt\right) \right] .
\end{equation}
Meanwhile the volume charge is increased:

\begin{equation}
Q_v(t)=-q\left( 1-\frac 1{\varepsilon _0}\right) +q\sum_j\exp \left(
-\upsilon _jt\right) \left[ A_j\cos \left( \omega _jt\right) +B_j\sin \left(
\omega _jt\right) \right] .
\end{equation}
The second boundary charge remains unchanged.

For $t>a/u$ we have

\begin{equation}
Q_{as}(t)=-\frac q2\left( 1-\frac 1{\varepsilon _0}\right) +q\sum_j\exp
\left( -\upsilon _j\tau \right) \left[ A_j\cos \left( \omega _j\tau \right)
+B_j\sin \left( \omega _j\tau \right) \right] ,
\end{equation}

\begin{eqnarray}
Q_v(t) &=&q\sum_j\exp \left( -\upsilon _jt\right) \left[ A_j\cos \left(
\omega _jt\right) +B_j\sin \left( \omega _jt\right) \right] - \\
&&-q\sum_j\exp \left( -\upsilon _j\tau \right) \left[ A_j\cos \left( \omega
_j\tau \right) +B_j\sin \left( \omega _j\tau \right) \right] .  \nonumber
\end{eqnarray}
The charge on the first boundary is given in this case by expression (30).

After the particle crosses the second boundary ($t>a/u$), the charge on the
first boundary decreases to its lower limit $(q/2)(1-1/\varepsilon _0)$. On
the second boundary the charge value increases and attains its maximum $%
-(q/2)(1-1/\varepsilon _0)$ when $t\gg a/u$. The wake charge in the volume
becomes equal to zero.

Thus it follows that after the particle crosses the first boundary, the
surface charge is transformed into the wake charge. The latter is
transformed again into the surface charge after the particle crosses the
second boundary.

We apply the results represented by expressions (20), (21) and (29) for the
induced charges in the model of diatomic cubic ionic crystal or polar
semiconductor, whose dielectric function is given by [27]

\begin{equation}
\varepsilon (\omega )=\varepsilon _\infty \frac{\omega _L^2-\omega
^2-i\upsilon \omega }{\omega _T^2-\omega ^2-i\upsilon \omega }.
\end{equation}
In this expression $\varepsilon _\infty $ is the optical frequency
dielectric function, $\omega _L$ and $\omega _T$ are the frequencies of the
longitudinal and transverse-optical vibration modes of infinite wavelength, $%
\upsilon $ is the damping rate, which we assume to be small ($\upsilon \ll
\omega _L$), and $\varepsilon _0$ is the static dielectric function, which
enters the theory through the Lyddane-Sachs-Teller relation, $\omega
_T^2\varepsilon _0=\omega _L^2\varepsilon _\infty $. Expression (34) implies
that its root $\varepsilon (\omega )=0$ has the following form:

\begin{equation}
\omega _j-i\upsilon _j=-\frac{i\upsilon }2+\Omega ,
\end{equation}
where $\Omega ^2=\omega _L^2-\upsilon ^2/4$. Substituting (34) and (35) into
expression (22), we find the coefficients determining the net induced
charges:

\begin{equation}
A_j\equiv A=\frac 1{\varepsilon _\infty }-\frac 1{\varepsilon
_0},~B_j=gA,~g=\frac \upsilon {2\Omega }.
\end{equation}
Figure 1 depicts the time dependence of $Q_{os}(t)$, $Q_{as}(t)$, and $%
Q_v(t) $ for the slab of $LiF$. The following values of the parameters were
taken for numerical calculations: $\varepsilon _\infty =1.96$ and $%
\varepsilon _0=9.01$ [27], $\upsilon /\omega _L=0.2$, and $\Omega a/u=15$.
We see that as the particle crosses the boundary, the surface and wake
charges oscillate with a frequency $\Omega $, although the net induced
charge remains equal to zero.

\section{Conclusion}

Let us briefly discuss the conditions in which the processes taking place at
the boundaries of the slab can be considered independent and the boundary
can be interpreted as that of a half-space, as was done by Gorbunov et al
[19].

We see from expressions (20), (21) and (29) that if the condition $%
a<u/\upsilon _j$ is met, the charge $-q(1-1/\varepsilon _0)$ has no time to
transform into the wake charge before the particle reaches the second
boundary. For this reason the transformation of a surface charge into a wake
charge and the transformation of the latter into a surface charge at the
second boundary are interrelated. When the particle crosses the second
boundary of the slab, near the boundary it excites electric-field
oscillations [3, 20, 26] whose phase is related to that of the oscillations
of the electric field near the first boundary. For the fields as the
boundaries to be completely independent, $a$ must exceed $u/\upsilon _j$. In
this case not only the amplitudes but also the phases of oscillations of the
electric fields at the boundaries are independent.

We would like to end our consideration with the following note. The
densities of induced surface charges (expressions (23)-(25)) and its
corresponding surface current densities may be considered as the sources of
transition radiation [27]. As follows from (24) and (25), the densities of
induced surface charges are determined by both electrostatic and
electromagnetic fields created by the moving particle. The total induced
surface charges, determined by (21) and (29), do not depend on the velocity
of light and, therefore, are determined only by electrostatic fields. Thus,
electromagnetic fields do not contribute into the total induced surface
charge. It seems that relativistic effects apparently reveal themselves in
the case when the thickness of the medium traversed is limited.

\newpage\ 

\begin{center}
{\bf Figure Caption}
\end{center}

Fig.1. The dynamics of the induced charge at the front boundary (full
curve), the rear boundary (broken curve), and in the volume (dotted curve)
for the slab of $LiF$. The following values of the parameters were taken for
numerical calculations: $\varepsilon _\infty =1.96$, $\varepsilon _0=9.01$, $%
\upsilon /\omega _L=0.2$, and $\Omega a/u=15$.

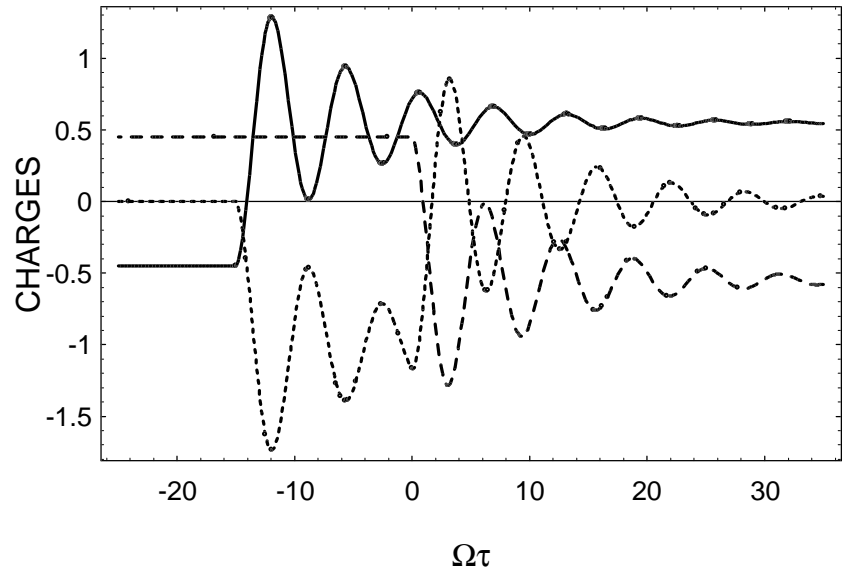
\end{document}